\documentclass[letterpaper,twocolumn,10pt]{article}
\usepackage{usenix2019_v3}

\usepackage{tikz}
\usetikzlibrary{calc}
\usepackage{amsmath}

\hypersetup{colorlinks=false}
\usepackage{hyperref}
\usepackage{tabularx}
\usepackage{listings}
\usepackage{algorithm}
\usepackage{algorithmic}
\usepackage{graphicx}
\usepackage{float}
\usepackage[bf,sc,small]{caption} 
\usepackage{subfig}
\usepackage{xcolor}
\usepackage{colortbl}
\usepackage{bytefield}
\usepackage[nolist]{acronym}
\usepackage{paralist}
\usepackage[textwidth=20mm,disable]{todonotes}

\usepackage{nopageno}
\usepackage{authblk}

\usepackage{enumitem}
\setlist[description]{leftmargin=0em,labelindent=0em}

\newcommand{\circlearound}[1]{\unitlength1ex\begin{picture}(2.0,2.0)%
\put(0.75,0.75){\circle{2.0}}\put(0.75,0.75){\makebox(0,0){#1}}\end{picture}}

\usepackage{titlesec}
\titlespacing*{\section} {0pt}{2.5ex plus 1ex minus .2ex}{2.3ex plus .2ex}
\titlespacing*{\subsection} {0pt}{2.25ex plus 1ex minus .2ex}{1.3ex plus .2ex}
\titlespacing*{\subsubsection} {0pt}{2.25ex plus 1ex minus .2ex}{1.3ex plus .2ex}

\begin{document}

\definecolor{MyGray}{gray}{0.80}

\begin{acronym}
 \acro{AES}{Advanced Encryption Standard}
 \acro{ASIC}{Application-Specific Integrated Circuit}
 \acro{BEL}{Basic Element}
 \acro{BRAM}{Block RAM}
 \acro{CLB}{Configurable Logic Block}
 \acro{CMOS}{Complementary Metal-Oxide-Semiconductor}
 \acro{CPA}{Correlation Power Analysis}
 \acro{DSP}{Digital Signal Processor}
 \acro{EM}{Electromagnetic}
 \acro{FF}{Flip-Flop}
 \acro{FPGA}{Field Programmable Gate Array}
 \acro{HD}{Hamming Distance}
 \acro{HDL}{Hardware Description Language}
 \acro{HW}{Hamming Weight}
 \acro{IC}{Integrated Circuit}
 \acro{LUT}{Look-Up-Table}
 \acro{PA}{Power Analysis}
 \acro{PRNG}{Pseudorandom Number Generator}
 \acro{RSA}{Rivest, Shamir and Adleman}
 \acro{SaM}{Square-and-Multiply}
 \acro{SNR}{Signal-to-Noise Ratio}
 \acro{SPA}{Simple Power Analysis}
 \acro{CBC}{Cipher Block Chaining}
 \acro{IP}{Intellectual Property}
 \acro{BBRAM}{battery-backed RAM}
 \acro{IFA}{Information Flow Analysis}
 \acro{GLIFT}{Gate-Level Information Flow Tracking}
 \acro{PUF}{Physical Unclonable Function}
 \acro{FSM}{Finite State Machine}
 \acro{IV}{initialization vector}
 \acro{RS}{Revision Select}
\end{acronym}
\date{}


\title{\Large \bf The Unpatchable Silicon: A Full Break of the Bitstream Encryption of\\Xilinx 7-Series FPGAs}


\author[*]{Maik Ender}
\author[*]{Amir Moradi}
\author[*$\dag$]{Christof Paar}
\affil[*]{Horst Goertz Institute for IT Security, Ruhr University Bochum, Germany}
\affil[$\dag$]{Max Planck Institute for Cyber Security and Privacy, Germany}

\renewcommand\Authands{ and }


\maketitle



\begin{abstract}

The security of FPGAs is a crucial topic, as any vulnerability within the hardware can have severe consequences, if they are used in a secure design.
Since FPGA designs are encoded in a bitstream, securing the bitstream is of the utmost importance.
Adversaries have many motivations to recover and manipulate the bitstream, including design cloning, IP theft, manipulation of the design, or design subversions e.g., through hardware Trojans. 
Given that FPGAs are often part of cyber-physical systems e.g., in aviation, medical, or industrial devices, this can even lead to physical harm.
Consequently, vendors have introduced bitstream encryption, offering authenticity and confidentiality.
Even though attacks against bitstream encryption have been proposed in the past, e.g., side-channel analysis and probing, these attacks require sophisticated equipment and considerable technical expertise.

In this paper, we introduce novel low-cost attacks against the Xilinx 7-Series (and Virtex-6) bitstream encryption, resulting in the total loss of authenticity and confidentiality.
We exploit a design flaw which piecewise leaks the decrypted bitstream. In the attack, the FPGA is used as a decryption oracle, while only access to a configuration interface is needed.
The attack does not require any sophisticated tools and, depending on the target system, can potentially be launched remotely. In addition to the attacks, we discuss several countermeasures.


\end{abstract}


\section{Introduction}
\label{sec:intro}





















Nowadays, \acp{FPGA} are common in consumer electronic devices, aerospace, financial computing, and military applications. 
Additionally, given the trend towards a connected world, data-driven practices, and artificial intelligence, FPGAs play a significant role as hardware platforms deployed in the cloud and in end devices. 
Hence, trust in the underlying platform for all these applications is vital.
Altera, who are (together with Xilinx) the FPGA market leader, was acquired by Intel in 2015. 

FPGAs are reprogrammable ICs, containing a repetitive logic area with a few hundred up to millions of reprogrammable gates. 
The bitstream configures this logic area; in analogy to software, the bitstream can be considered the `binary code' of the FPGA. 
On SRAM-based FPGAs, which are the dominant type of FPGA in use today,  the bitstream is stored on an external non-volatile memory and loaded into the FPGA during power-up.

In order to protect the bitstream against malicious actors, its confidentiality and authenticity must be assured. 
If an attacker has access to the bitstream and breaks its confidentiality, he can reverse-engineer the design, clone intellectual property, or gather information for subsequent attacks e.g., by finding cryptographic keys or other design aspects of a system. 
If the adversary succeeds in violating the bitstream authenticity, he can then change the functionality, implant hardware Trojans, or even physically destroy the system in which the FPGA is embedded by using configuration outside the specifications. 
These problems are particularly relevant since access to bitstream is often effortlessly possible due to the fact that, for the vast majority of devices, it resides in the in \emph{external} non-volatile memory, e.g., flash chips. 
This memory can often either be read out directly, or the adversary wiretaps the FPGA's configuration bus during power-up. 
Alternatively, a microcontroller can be used to configure the FPGA, and consequently, the microcontroller's firmware includes the bitstream. 
When the adversary gains access to the microcontroller, he also gains access to the configuration interface and the bitstream. 
Thus, if the microcontroller is connected to a network, remotely attacking the FPGA becomes possible.

In order to protect the design, the major FPGA vendors introduced bitstream encryption around the turn of the millennium, a technique which nowadays is available in most mainstream devices~\cite{AlteraBSEnc, ug7Config}. 
In this paper, we investigate the security of the Xilinx 7-Series and Virtex-6 bitstream encryption. 
On these devices, the bitstream encryption provides authenticity by using an SHA-256 based HMAC and also provides confidentiality by using CBC-AES-256 for encryption.
By our attack, we can circumvent the bitstream encryption and decrypt an assumedly secure bitstream on all Xilinx 7-Series devices completely and on the Virtex-6 devices partially.
Additionally, we are also able to manipulate the bitstream by adjusting the HMAC.
Out attack setting in general is the same one as commonly encountered in mainstream practice: The adversary only needs access to the configuration interface of a fielded FPGA. In this setting, the secret decryption key has already been loaded into the FPGA, e.g., after device manufacturing, the key is stored in internal \ac{BBRAM} or eFUSEs. As will be shown later, the adversary uses the FPGA with the stored key as an oracle to decrypt the bitstream.

According to recent business reports, Xilinx shares 50\% of the FPGA market~\cite{XilinxMarketShare}.  
Also evident by Xilinx's annual report in 2018~\cite{Xilinx18Form10k}, around 35\% of their current revenue originates from the 7-Series
(meanwhile, Virtex-6 devices are not stated independently in this report, but are veiled in the 50\% revenue of all old generations).
Thus, the 7-Series and Virtex-6 devices are a popular choice for a variety of FPGA designs, many of which are mission- or safety-critical. 
Besides, we note that similar to many other digital hardware devices, FPGAs have a lifespan of decades.
Replacing legacy systems or using high-performance products therefore might turn out to be a costly and cumbersome undertaking. 
However, Xilinx's new UltraScale and UltraScale+ devices, which are the new (high-end) series and slowly replace the old ones, are not affected by our attack.

In this paper, we introduce two novel attacks against this Xilinx 7-Series bitstream encryption, which result in a total loss of authenticity and confidentiality.
Furthermore, we discuss the implications of these attacks and suggest potential countermeasures. 
While our attacks chiefly target the Xilinx 7-Series, Virtex-6 devices are also vulnerable to our attack with the limitation that the first two bits of every 32-bit word are missing in the recovery process.

We communicated our findings to Xilinx in a vulnerability disclosure on 24 September 2019 and started cooperating on the issue:
Xilinx quickly confirmed the vulnerability on 25 September and that there is no patch possible without changing the silicon. 
Coinciding with the publication of this paper, Xilinx plans to publish a design advisory that informs their customers of this vulnerability.

The paper is structured as follows: First, we give an executive summary of the attack.
Then, we introduce the necessary background and related work in Section~\ref{sec:background}.
In Section~\ref{sec:attack}, we introduce the attack with all details, whereupon we validate the attack by a case study in Section~\ref{sec:caseStudys}.
A discussion about the findings and countermeasures is given in Section~\ref{sec:countermeasures}.
We conclude the paper in Section~\ref{sec:conclusion}.

\subsection{The Attack at a Glance}
A small configuration engine loads the bitstream into the FPGA and continuously reflects the FPGA's state in status registers.
If the bitstream encryption is activated, the configuration engine prohibits the readout of a bitstream.
Usually, if the bitstream encryption is disabled, this readout function is legitimately used for debugging the FPGA and its design.

In our attack, we manipulate the encrypted bitstream to redirect its (decrypted) content from the fabric to a configuration register.
We then read out this configuration register, which holds the unencrypted bitstream data;
the readout of the configuration register is not prevented even in the presence of an encrypted bitstream anyway.

For that purpose, we use the MultiBoot address register WBSTAR.
This MultiBoot feature enables the FPGA to boot from a different memory address in order to update the FPGA safely, boot with different functionality or boot from a fallback bitstream with a working design. 
The MultiBoot feature uses the content of the WBSTAR register as the boot address in the attached non-volatile memory.
Hence, the register is not cleared during a reset. 
We now manipulate the encrypted bitstream to write a single 32-bit word which is part of the encrypted bitstream to the WBSTAR register in \emph{decrypted form}.
The bitstream's manipulation exploits the malleability of the CBC mode of operation to alter the command in the bitstream which writes data to the WBSTAR configuration register.
After the configuration with the encrypted bitstream, the FPGA resets, since it detects an invalid HMAC.
We use the WBSTAR configuration register for the readout, because the reset procedure does not clear it.
After the reset, we finally use a second bitstream to readout the WBSTAR register to uncover the decrypted bitstream word by word.
In summary, the FPGA, if loaded with the encryption key, decrypts the encrypted bitstream and writes it for the attacker to the readable configuration register.
Hence, the FPGA is used as a decryption oracle. 
The fact that only single 32-bit words can be uncovered in each iteration determines the duration of decrypting a whole bitstream:
In our experiments, we are able to uncover a complete Kintex-7 XC7K160T bitstream in 3~hours and 42~minutes, for instance.

For the second attack, we can break the authenticity of the bitstream encryption.
The attacker can use the decryption oracle to encrypt arbitrary messages due to the underlying CBC mode.
They can build the CBC chain starting with the last block.
For that, they encrypt a random message, uses the CBC malleability, and calculates the ciphertext block to turn the plaintext into the intended value.
The attacker repeats this process until the whole bitstream is encrypted.
Since the HMAC key is stored in the encrypted bitstream and is not verified, the attacker can manipulate the HMAC tag as well.
Thus, the attacker can craft legitimate encrypted bitstreams, which are correctly validated.



\section{Background}

\label{sec:background}
In this section, we introduce the background on FPGAs, give an overview of attacks already mounted on bitstream encryption schemes, and lastly, introduce the bitstream format of the Xilinx 7-Series. 

\subsection{FPGAs}
\aclp{FPGA} are reconfigurable devices.
They consist, in essence, of an array of configurable logic cells, also known as fabric.
The main elements of the fabric are small configurable logic cells, flip-flops, and a configurable routing.
Only if the user programs the FPGA, it contains the functional logic of the design.
The most significant advantage of FPGAs over ASICSs is their reprogrammability, i.e., the ability to configure an FPGA arbitrarily.

All configuration information is contained in  the bitstream, which specifies all details of the digital design.
In SRAM-based FPGAs, it has to be stored on an external non-volatile memory chip.
For programming the bitstream, the FPGA has different interfaces, e.g., SelectMAP, JTAG, ICAP2, Serial, or SPI/BPI.
The difference between these interfaces are mostly their protocol, bus width, and direction of programming, i.e., the SPI interface independently reads from non-volatile memory, while the SelectMAP or JTAG can be triggered from another device and the ICAP2 is an internal port inside the fabric.
Additionally, the SelectMAP, JTAG, and ICAP2 interfaces have a back-channel, i.e., they can read out debug information from the FPGA. 
This readout enables the user to download the configured design, e.g., extract the bitstream from the FPGA and check if anything was configured correctly or use the flip-flop content for advanced design debugging.
Similarly, the user can read out the configuration and status registers from the FPGA.

The bitstream encryption feature protects the bitstream by providing confidentiality and authenticity.
The encryption key is stored in either a \ac{BBRAM} or eFUSEs and is programmed via JTAG only. 
\todo[inline]{figure oder noch mehr schreiben?}
\todo[inline]{AM: I would go for a figure generally show how the bitstream encryption works.}
When the bitstream encryption is enabled, the readout of the bitstream described above is blocked on all external ports.
Otherwise, an attacker would be able to read out the decrypted design information.
Hence, a readout from the external ports returns \texttt{null} values when the bitstream encryption is used.
Only via the internal ICAP2 interface, it is possible to read out the encrypted bitstream.
However, the ICAP2 interface is usually not connected to the outside world or should be protected.
An additional security mechanism is that the entire FPGA must be reset to load a new design when the bitstream encryption has been enabled.
%

\subsection{Bitstream-Based Attacks}
The consequence of our attack is the total loss of the bitstream's authenticity and confidentiality.
Even when losing one of them, attacks against the system become possible~\cite{trimberger2014fpga}.
A recent example is the Thrangrycat attack of Kataria et al., which targets the FPGA-based root of trust in Cisco routers~\cite{kataria2019defeating}. 
In this section, we elaborate the following attacks and their implications: cloning, reverse engineering, tampering, spoofing, and physically harming of FPGAs and their design.
Besides that, the general security of FPGAs is a well-studied topic in the literature~\cite{skorobogatov2012breakthrough, herklotz2019finding, skorobogatov2019hardware, moradi2011vulnerability, trimberger2014fpga, lohrke2018key}, which will be discussed mainly in the next chapter.

Without bitstream confidentiality, the design can easily be cloned and counterfeit products can be built.
Thus, overproduction is considered a considerably higher threat in the case of FPGAs compared to ASIC-based products.
A bitstream without confidentiality also allows that the design can be reverse engineered to gain knowledge about the \ac{IP} used, mount attacks on the application, or prepare the injection of hardware Trojans.
Hardware Trojans and other manipulation attacks are based on tampering with the bitstream.
Thus, an adversary has also to circumvent the bitstream's authenticity.
 We note that manipulations allow the attacker also to circumvent other security mechanisms in the design or leak data within the design, e.g., cryptographic keys.
 Moreover, an attacker might be able to physically destroy the system in which it is embedded by changing parameters, akin to the Stuxnet attack (which was allegedly software-based, however) \cite{langner2011stuxnet}.

When spoofing the bitstream, the attacker replaces the bitstream rather than changing the already existing one, i.e., the attacker creates his own bitstream.
Thus, no reverse engineering of the existing bitstream is needed.

Another bitstream-based attack vector is to physically destroy the FPGA by configuration outside the specifications, i.e., by implementing short circuits on the FPGA.
Physical harming the FPGA through its fabric might not be necessary for an attacker with access to the hardware, as they can anyhow destroy it.
However, an attacker with only remote access to the bitstream will be capable of physically harming the inside of the FPGA.
Such physical attacks can be viewed as severe denial of service attacks.

Thus, almost all vendors realized means to secure the bitstream of an FPGA.
First, they block the readback of the bitstream from debugging interfaces.
Second, they have developed  bitstream encryption schemes.
The bitstream encryption should provide authenticity and confidentiality, as the confidentiality protects the bitstream against cloning, reverse engineering, and tampering~\cite{trimberger2014fpga}, while the authenticity is needed to avoid loading an untrusted bitstream and prevent tampering, spoofing, and physical harm attacks.
As otherwise, the attacker could run a modified bitstream on the device. 
Thus, the authenticity of the bitstream is as essential as its confidentiality~\cite{drimer2007authentication}.

\subsection{Related Works}
Several attacks against  bitstream encryption have been proposed in the literature.
In 2012, Skorobogatov and Woods found a bug (which might be a backdoor) to circumvent the bitstream encryption of an Actel/Microsemi ProA-SIC3 A3P250 FPGA~\cite{skorobogatov2012breakthrough}.
They found a bug in the JTAG instruction set to read out the bitstream even when the bitstream encryption is enabled.

Already in 2011, Moradi et al. attacked the Xilinx bitstream encryption with power side-channel attacks~\cite{moradi2011vulnerability}.
Subsequently, in 2014, Altera FPGAs have been targeted using side-channels~\cite{swierczynski2014physical} as well.
After measuring the power consumption of the device, the attacker uses statistical methods and a power-model of the cipher to compute the key.
Often GPUs are used to compute the key from an ample search space in a reasonable time frame.
Furthermore, the PCB hosting the FPGA often needs to be modified to allow monitoring the power side-channel.
This requirement is relaxed by measuring the electro-magnetic side channel instead of the power side channel~\cite{moradi2016improved} but comes at an increase in the measurement cost and complexity. 
For example, Moradi et al. used an oscilloscope at a sampling rate of 5\,GS/s and bandwidth of 1.5\,GHz to capture EM signals.
Since its introduction, these side-channel attacks have become a general thread to bitstream encryption schemes, which led to improved countermeasures in recent FPGA series.
Nevertheless, the general knowledge on side-channel attacks has improved during the last decades, and the number of companies and research institutes active in the field has grown.
Although this increases the feasibility of such attacks, the adversary requires a minimum set of equipment to be able to measure side-channel leakages with adequate quality.

Tajik et al. introduced an attack using optical contactless probing in 2017~\cite{tajik2017power}. 
In a nutshell, a near-infrared light source is focused on the backside of the silicon, i.e., directly on the transistors.
The hereby used near-infrared light source is transparent to the substrate.
Thus, it directly reaches the transistors.
The transistors then reflect the emitted light depending on their load.
Consequently, a detector can distinguish between a transistor in an opened or closed state.
The authors used this technique to attack the bitstream encryption of a Xilinx Kintex-7 FPGA successfully. They observed the bitstream configuration engine and identified the bus transmitting the plaintext bits after the encryption. 
Hence, they used the FPGA as an oracle, as well. 
Nevertheless, this attack requires expensive electro-optical probing equipment.

Similarly, thermal laser stimulation attacks~\cite{lohrke2018key} uses laser beams to introduce localized heating, which changes the used current. The current changes can then be linked to the stored key in the BBRAM to extract the encryption key.

Lately, security researchers at F-Secure points out two design flaws in the encrypt-only boot mode of Zynq UltraScale+ MPSoC devices~\cite{CVEXilinxUSHack}, which compromise the processing unit (ARM core) in the SoC design. 
The researchers shows that the header of the first stage boot loader is not checked, which encodes the boot start address of the processing unit (ARM core).
Changing the address can lead to arbitrary code execution using a return-oriented programming attack.
Nevertheless, the attack is mountable in the encrypt only boot mode solely.
Hence, it can be mitigated, as recommended by Xilinx before, by using system level protections or the Hardware Root of Trust boot mode, which uses RSA signatures to authenticate the boot header. 

In summary, the known attacks to the Xilinx bitstream encryption on 7-Series devices are all physical in nature (side-channel analysis, optical contactless probing), and are mostly costly in terms of equipment, time, and technical expertise.
Plus, they need physical access to the FPGA.
In contrast, our attack requires only access to a JTAG or SelectMAP interface, which is often available through the debugging nature of the JTAG interface or may be even available via a remote channel.


\subsection{Bitstream Format}
The Xilinx 7-Series bitstream format contains a header and the configuration for the fabric.
While most of the header is documented in~\cite{ug7Config}, the fabric configuration is not made public by the vendor.
However, several papers show strategies to document the bitstream format~\cite{ender2019insights, prjxray, rannaud2008bitstream,debit, max5, JBITS, DingWZZ13}, as the fabric configuration data is the netlist, in a different format, of the loaded design.
Hence, the fabric data format is essential for reverse engineering of the design, to find Trojan horses, to build open-source tool-chains, or to formally verify the bitstream coming from the vendors' tools.

\begin{figure}
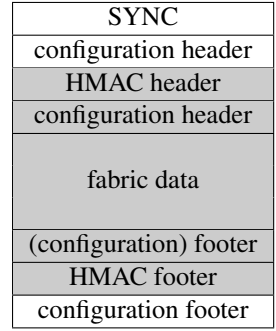

 \begin{center}
\begin{tabular}{|c|}\hline
SYNC \\\hline
configuration header \\\hline
\rowcolor{MyGray} 
HMAC header \\\hline
\rowcolor{MyGray} 
configuration header \\\hline
\rowcolor{MyGray} \\
\rowcolor{MyGray} 
fabric data\\
\rowcolor{MyGray} 
\\\hline
\rowcolor{MyGray} 
(configuration) footer\\\hline
\rowcolor{MyGray} 
HMAC footer\\\hline
configuration footer \\\hline
\end{tabular}
\caption{Bitstream structure overview (shaded parts are encrypted)~\cite{ug7Config, trimberger2014fpga}.}
\label{fig:bitstreamOverview}
\end{center}
\end{figure}

\begin{figure*}
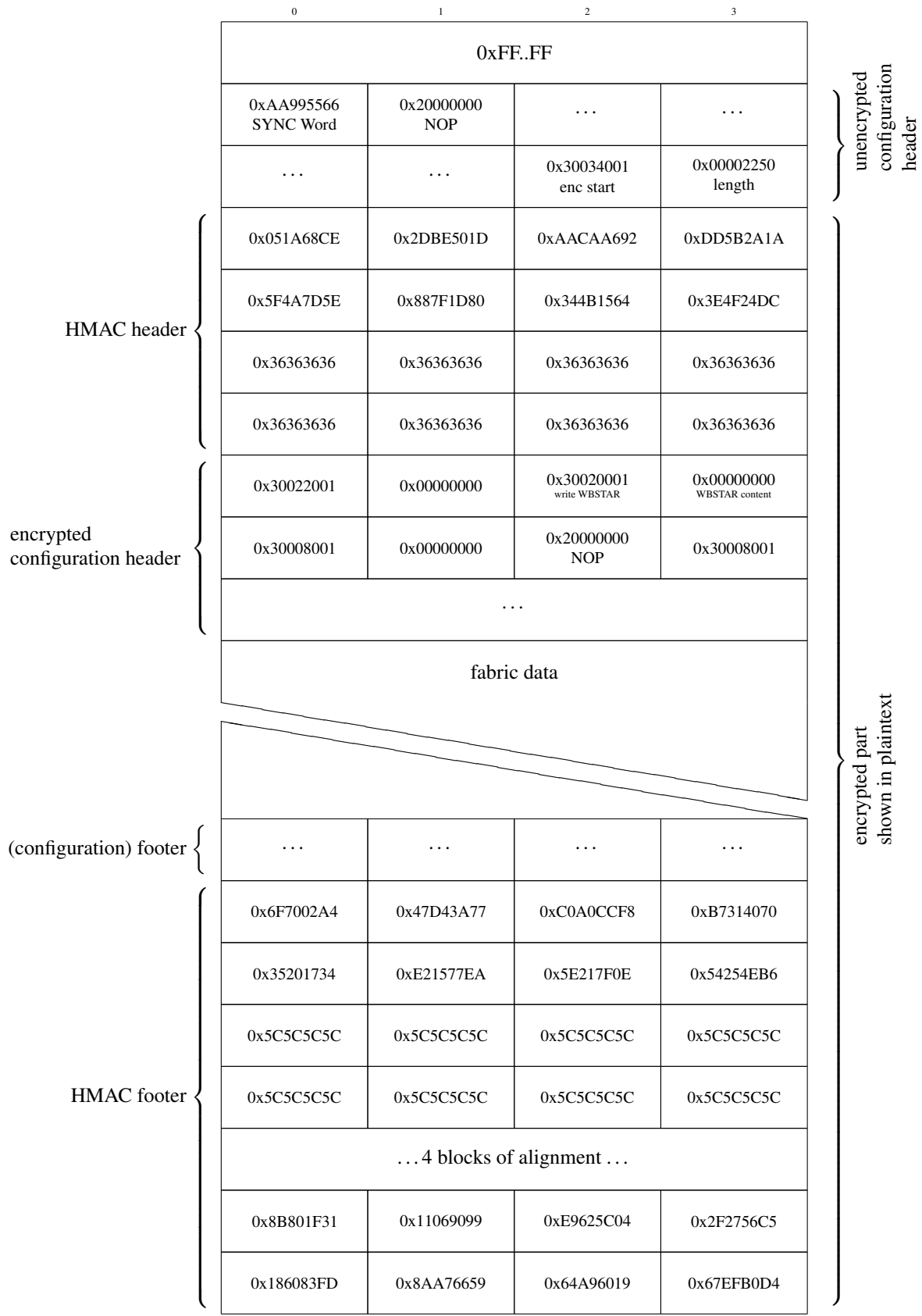

\centering
\begin{bytefield}[bitwidth=7.1em,bitheight=3em]{4}
	\small
	\bitheader{0-3} \\
	\bitbox{4}{0xFF..FF} \\
	\begin{rightwordgroup}{\rotatebox{90}{\parbox{1cm}{unencrypted\\configuration\\header}}}
		\bitbox{1}{\footnotesize 0xAA995566\\{SYNC Word}} & \bitbox{1}{\footnotesize 0x20000000\\NOP} & \bitbox{1}{\dots} & \bitbox{1}{\dots} \\
	
		\bitbox{1}{\dots} & \bitbox{1}{\dots} & \bitbox{1}{\footnotesize 0x30034001\\enc start} &\bitbox{1}{\footnotesize 0x00002250\\length}
	\end{rightwordgroup} \\
	\begin{rightwordgroup}{\rotatebox{90}{\parbox{3cm}{encrypted part\\shown in plaintext}}}
		\begin{leftwordgroup}{HMAC header}
			\bitbox{1}{\footnotesize 0x051A68CE} & \bitbox{1}{\footnotesize 0x2DBE501D} & \bitbox{1}{\footnotesize 0xAACAA692} & \bitbox{1}{\footnotesize 0xDD5B2A1A} \\
			\bitbox{1}{\footnotesize 0x5F4A7D5E} & \bitbox{1}{\footnotesize 0x887F1D80} & \bitbox{1}{\footnotesize 0x344B1564} & \bitbox{1}{\footnotesize 0x3E4F24DC} \\
			\bitbox{1}{\footnotesize 0x36363636} & \bitbox{1}{\footnotesize 0x36363636} & \bitbox{1}{\footnotesize 0x36363636} & \bitbox{1}{\footnotesize 0x36363636} \\
			\bitbox{1}{\footnotesize 0x36363636} & \bitbox{1}{\footnotesize 0x36363636} & \bitbox{1}{\footnotesize 0x36363636} & \bitbox{1}{\footnotesize 0x36363636} 
		\end{leftwordgroup}\\
		
		\begin{leftwordgroup}{\parbox{3cm}{encrypted\\configuration header}}
			\bitbox{1}{\footnotesize 0x30022001} & \bitbox{1}{\footnotesize 0x00000000} & 
			\bitbox{1}{\footnotesize 0x30020001\\\tiny write WBSTAR} & \bitbox{1}{\footnotesize 0x00000000\\\tiny WBSTAR content} \\
			
			\bitbox{1}{\footnotesize 0x30008001} & \bitbox{1}{\footnotesize 0x00000000} & 
			\bitbox{1}{\footnotesize 0x20000000\\NOP} & \bitbox{1}{\footnotesize 0x30008001} \\
			
			\bitbox{4}{\dots}
			
		\end{leftwordgroup}\\

		\wordbox[lrt]{1}{fabric data} \\
		\skippedwords \\

		\begin{leftwordgroup}{(configuration) footer}
			\bitbox{1}{\dots} &\bitbox{1}{\dots} &\bitbox{1}{\dots} &\bitbox{1}{\dots} & 
		\end{leftwordgroup}\\
		\begin{leftwordgroup}{HMAC footer}
			\bitbox{1}{\footnotesize 0x6F7002A4} & \bitbox{1}{\footnotesize 0x47D43A77} & \bitbox{1}{\footnotesize 0xC0A0CCF8} & \bitbox{1}{\footnotesize 0xB7314070} \\
			\bitbox{1}{\footnotesize 0x35201734} & \bitbox{1}{\footnotesize 0xE21577EA} & \bitbox{1}{\footnotesize 0x5E217F0E} & \bitbox{1}{\footnotesize 0x54254EB6} \\
			\bitbox{1}{\footnotesize 0x5C5C5C5C} & \bitbox{1}{\footnotesize 0x5C5C5C5C} & \bitbox{1}{\footnotesize 0x5C5C5C5C} & \bitbox{1}{\footnotesize 0x5C5C5C5C} \\
			\bitbox{1}{\footnotesize 0x5C5C5C5C} & \bitbox{1}{\footnotesize 0x5C5C5C5C} & \bitbox{1}{\footnotesize 0x5C5C5C5C} & \bitbox{1}{\footnotesize 0x5C5C5C5C} \\
			\bitbox{4}{\dots 4 blocks of alignment \dots} \\
			
			\bitbox{1}{\footnotesize 0x8B801F31} & \bitbox{1}{\footnotesize 0x11069099} & \bitbox{1}{\footnotesize 0xE9625C04} & \bitbox{1}{\footnotesize 0x2F2756C5} \\
			\bitbox{1}{\footnotesize 0x186083FD} & \bitbox{1}{\footnotesize 0x8AA76659} & \bitbox{1}{\footnotesize 0x64A96019} & \bitbox{1}{\footnotesize 0x67EFB0D4} 
		\end{leftwordgroup}
%
%
%
%

	\end{rightwordgroup}
	
\end{bytefield}

	\caption{Bitstream structure }
	\label{figure:bitstream_structure}
\end{figure*}

Figure~\ref{fig:bitstreamOverview} shows an overview of the 7-Series encrypted bitstream structure.  
Later in Figure~\ref{figure:bitstream_structure}, we discuss the bitstream format in detail.
The bitstream starts with a SYNC word, which is followed by a configuration header. 
In the header, the CBC IV is configured and the length of the following encrypted part is given.
After the header, the encrypted part follows, which is shaded in Figure~\ref{fig:bitstreamOverview}. 
First, in the HMAC header, the HMAC key (ipad) is set, which immediately starts the HMAC calculation.  
Then, a secondary header configures the remaining settings.
A large blob is followed to configure the fabric. 
A footer concludes the configuration, which is also used for alignment.
The encrypted part ends with the HMAC footer, which contains the HMAC opad, and the HMAC tag, with which the encrypted part can be validated.
A global footer concludes the bitstream as well as starting the FPGA's fabric.

In detail, the bitstream of 7-Series devices is organized in packages of 32-bit words.
There are two types of packages, while the type 1 package is displayed in Table~\ref{tab:type1Package}.
Type 1 packages contain an opcode (nop, read, write), a register address, and the word count of the read or written data.
If a package writes any content to a register, the data (in multiples of 32-bit words) is attached directly after the package.
The type 2 package is an extension of the type 1 with a larger address field to write a large amount of data, e.g., the fabric data.

\begin{table*}
 \begin{center}
\begin{tabular}{|c|c|c|c|c|}\hline
\textbf{Header Type} & \textbf{Opcode} & \textbf{Register Address} & \textbf{Reserved} & \textbf{Word Count}\\\hline
[31:29] & [28:27] & [26:13] & [12:11] & [10:0]\\\hline
001 & xx & RRRRRRRRRxxxxx & RR & xxxxxxxxxxx\\\hline
\end{tabular}
\caption{Type 1 package header format. ``R'' are reserved bits and ``x'' are the actually used bits~\cite{ug7Config}.}
\label{tab:type1Package}
\end{center}
\end{table*}

There are 20 documented registers, which organize the configuration of the FPGA.
For example, there is a CRC register verifying the checksum of the bitstream or multiple status registers to monitor the boot process. 
The interested reader is referred to the documentation of the bitstream header format~\cite{ug7Config}. 


In more detail, Figure~\ref{figure:bitstream_structure} shows the structure of an encrypted bitstream.
The configuration logic ignores the beginning of the bitstream until the sync word 0xAA995566 is transmitted.
The following unencrypted header configures only the decryption engine, i.e., turns it on and sets the CBC IV.
With writing the length of the encrypted part in the configuration register 
\footnote{0x30034001 is the command (type 1 package), where 0x3 is a write command, 0x34 determine the written register, and 0x1 the length of data written to the register.
The following word 0x00002250 is the data written to that register 0x34 (see Table~\ref{tab:type1Package}).}, 
the bitstream encryption engine is turned on, and only encrypted data follows.
Note that we show all bytes in the encrypted part in plaintext as the FPGA configuration logic would see it after the decryption. 
However, the attacker would observe arbitrary encrypted data only.
The first 4 AES blocks, i.e., $4 \times 128 \textrm{ bits} = 512 \textrm{ bits}$, correspond to the HMAC header. 
It includes the HMAC key xored with the ipad (256-bits) and the ipad value itself~\cite{pub2002198}.
Following the HMAC header, the already started configuration header is completed by issuing commands to prepare the configuration engine, e.g., the WBSTAR register is configured. 
The word following the command to write to the WBSTAR register is the content written to the register, i.e., 0x00000000.
The configuration header completes with a command to write the fabric's data, which follows afterwards and is the longest part of the bitstream.

At the end of the encrypted part, the (configuration) footer and the HMAC footer is attached. 
The configuration footer can contain commands to configure the engine and an uninterpreted part, which is used to align the plaintext to a multiple of 512-bits since the HMAC operates on multiples of 512-bits. 
The HMAC footer contains the HMAC's key XORed with the opad, the opad itself, and the HMAC tag. 
The HMAC authenticates all encrypted content since the HMAC computation starts directly after the first HMAC header and ends right before the HMAC footer. 
Since the configuration engine processes all HMAC related calculations on the decrypted bitstream, the MAC-then-encrypt scheme is used.
Lastly, the general footer (in plain) ends the whole bitstream. 
Since we can ignore it for the attack, it is not shown in Figure~\ref{figure:bitstream_structure}.

%
%
%
%
%
%

\section{Attacking Xilinx Bitstream Encryption}
\label{sec:attack}
This section presents the adversary model, the malleability of the bitstream encryption, and gives an introduction on how to forge arbitrary bitstreams. We use the following notation: A word is 32-bit long, a block is 128-bit long (AES-256 operation) and a chunk is 512-bit long (one SHA-256 input).

\subsection{Adversary Model}

%

Generally, the adversary can be anyone who has access to the JTAG or SelectMAP configuration interface, even remotely, and to the encrypted bitstream of the device under attack.
In contrast to side-channel and probing attacks against bitstream encryption, no adequate equipment nor expertise in electronic measurements is needed. 
The requirements for our adversary model are as follows:

\begin{description}[labelindent=0cm]
\item[Configuration Interfaces] The attacker needs to have access to the SelectMAP or JTAG interface which allows a debug readout as well as the configuration of encrypted bitstreams. 
For example, the attacker can gain access to a configuration interface locally, if they have physical access.
Note that only the JTAG interface can be used to load the AES key. 
Thus, a JTAG interface must be present on the PCB, if the \ac{BBRAM} key storage is used. 
If the BBRAM is not used, the eFUSEs can be burned during provisioning on a different PCB, then used on the production PCB.
Hence, the JTAG interface might not be present on the production PCB, if eFUSEs are used, so that another configuration interface is used, like SelectMAP.

A microcontroller is often used  in addition to the FPGA and configures it. \todo[inline]{, often via the SelectMAP or JTAG interface.}
Thus, the attack can be conducted from the connected microcontroller, if it is connected to the FPGA's SelectMAP or JTAG interface.
It is even possible to conduct the attack remotely if the microcontroller is connected to a network, and because of that the adversary can gain access to the microcontroller via the remote channel, e.g., by installing a rootkit, as demonstrated by the latest Thrangrycat attack on Cisco routers~\cite{Thrangrycat}.

 \item[Bitstream Access] The adversary also needs to have access to the attacked encrypted bitstream, which they can through several methods.
In the case of physical access, they can wiretap the configuration bus during power-up or directly read out the non-volatile memory in which the encrypted bitstream is stored. 
Without physical access, however, they can extract it from the (microcontroller) firmware which configures the FPGA or download the firmware from a remote update service e.g., via a website.

 \item[Access Level] 
 \ 
 A \emph{remote attack} is possible only if the attacker has remote access to a configuration interface and the encrypted bitstream.
 
 A \emph{local attack} is possible otherwise, e.g., if the attacker has local access to a configuration interface and can obtain the encrypted bitstream.

\item[Key Loaded] The AES key must already be loaded onto the FPGA, which is always the case for a system already in use and may be the case after the provisioning by the system manufacturer.

\item[Known-plaintext] The attacker needs only limited knowledge about the plaintext of the encrypted bitstream.
Specifically, they need to know about a single 32-bit word in the encrypted bitstream header, since a single word is altered in the attack only. 
The bitstream generation in Vivado is deterministic, i.e., the commands in the encrypted header are the same among different bitstreams and change only in their configuration content. 
Thus, the adversary can predict the plaintext in the encrypted header, e.g., they know the position of the write WBSTAR command. 
Note that any other command can be used as long as the attacker knows the plaintext.
If a defender would change the encrypted bitstream header, e.g., randomize it, an attacker can make assumptions, as there is a limited set of valid packages e.g., package construction (Table~\ref{tab:type1Package}), valid commands, and meaningful content.
Therefore, the attacker could brute-force the encrypted bitstream to gain knowledge over the plaintext;
however, only a single package in the header needs to be brute-forced for this attack, so it would become more difficult, but not infeasible.

\item[Used Devices] The design under attack is any Xilinx 7-Series device or a Vertix-6 device with slight limitations (see Section~\ref{sec:caseStudys}). \todo[inline]{do we need this here, see reviewers?}

\end{description}

\subsection{CBC Malleability}
\label{sec:cbc_malleability}

\begin{figure*}
\centering

\begin{tikzpicture}

        \begin{scope}
            \def\x{0}
            \node (f\x) at ($\x*(2.5cm,0)$) [minimum size=1.25cm,rounded corners=1ex,fill=blue!20,draw] {{\sc Dec}};

            \node (c\x) [below of=f\x, node distance=2.5cm] {$P_\x$};

            \node (k\x) [left of=f\x, node distance=1.5cm] {$k$};

            \node (p\x) [below of=f\x, node distance=1.5cm, circle, draw] {};

            \node (m\x) [above of=f\x, node distance=1.5cm] {$C_\x$};

            \draw[-] (p\x.north) -- (p\x.south);

            \draw[-] (p\x.east) -- (p\x.west);

            \draw[-latex] (m\x) -- (f\x);

            \draw[-latex] (f\x) -- (p\x);

            \draw[-latex] (k\x) -- (f\x);

            \draw[-latex] (p\x) -- (c\x);

        \end{scope}

        \begin{scope}
            \def\x{1}
            \node (f\x) at ($\x*(2.5cm,0)$) [minimum size=1.25cm,rounded corners=1ex,fill=blue!20,draw] {{\sc Dec}};

            \node (c\x) [below of=f\x, node distance=2.5cm] {\color{red} $P_i'$};

            \node (k\x) [left of=f\x, node distance=1.5cm] {$k$};

            \node (p\x) [below of=f\x, node distance=1.5cm, circle, draw] {};

            \node (m\x) [above of=f\x, node distance=1.5cm] {$C_\x \color{red} \oplus \Delta$};

            \draw[-] (p\x.north) -- (p\x.south);

            \draw[-] (p\x.east) -- (p\x.west);

            \draw[-latex] (m\x) -- (f\x);

            \draw[-latex] (f\x) -- (p\x);

            \draw[-latex] (k\x) -- (f\x);

            \draw[-latex] (p\x) -- (c\x);

        \end{scope}

        \begin{scope}
            \def\x{2}
            \node (f\x) at ($\x*(2.5cm,0)$) [minimum size=1.25cm,rounded corners=1ex,fill=blue!20,draw] {{\sc Dec}};

            \node (c\x) [below of=f\x, node distance=2.5cm] {$P_\x \color{red} \oplus \Delta$};

            \node (k\x) [left of=f\x, node distance=1.5cm] {$k$};

            \node (p\x) [below of=f\x, node distance=1.5cm, circle, draw] {};

            \node (m\x) [above of=f\x, node distance=1.5cm] {$C_\x$};

            \draw[-] (p\x.north) -- (p\x.south);

            \draw[-] (p\x.east) -- (p\x.west);

            \draw[-latex] (m\x) -- (f\x);

            \draw[-latex] (f\x) -- (p\x);

            \draw[-latex] (k\x) -- (f\x);

            \draw[-latex] (p\x) -- (c\x);

        \end{scope}

        \node (iv) [left of=p0, node distance=1.5cm] {$IV$};

        \draw[-latex] (iv) -- (p0);

        \foreach \x in {0, 1} {

            \draw[-latex] ($(m\x) - (0,0.5cm)$) -| +(0.8cm,-2.5cm) -- ($(p\x) + (2.4cm,0)$);

        \begin{scope}

            \node at (6.4,0) {$\cdots\cdots$};

        \end{scope}

        \begin{scope}

            \node (f) at (9.6cm,0) [minimum size=1.25cm,rounded corners=1ex,fill=blue!20,draw] {{\sc Dec}};

            \node (c) [below of=f, node distance=2.5cm] {$P_n$};

            \node (k) [left of=f, node distance=1.5cm] {$k$};

            \node (p) [below of=f, node distance=1.5cm, circle, draw] {};

            \node (m) [above of=f, node distance=1.5cm] {$C_n$};

            \draw[-] (p.north) -- (p.south);

            \draw[-] (p.east) -- (p.west);

            \draw[-latex] (m) -- (f);

            \draw[-latex] (f) -- (p);

            \draw[-latex] (k) -- (f);

            \draw[-latex] (p) -- (c);

            \draw[-] ($(m) - (2.5,0.5cm)$) -- ($(m) - (1.7,0.5cm)$);

            \draw[-latex] ($(m) - (1.7,0.5cm)$) |- + (0cm,-2.5cm) -- (p);

        \end{scope}

        }

    \end{tikzpicture}
 \caption{CBC malleability during  decryption (figure based on~\cite{TikZ:for:Cryptographers}}
 \label{figure:cbc_malleability}
\end{figure*}
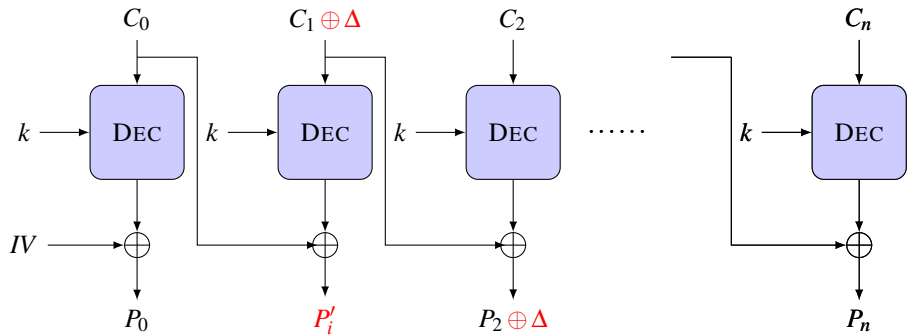

Xilinx uses the \acf{CBC} mode with AES as the underlying cipher for bitstream encryption.
Hence, the blocks are 128 bits wide. 
In \ac{CBC} mode, each ciphertext block $C_i$ is XORed with the next plaintext block $P_{i+1}$ prior to encryption. 
An advantage of the \ac{CBC} mode is that it encrypts probabilistically if a nonce is used as an \ac{IV}, which is a desirable security feature. However,  the \ac{CBC} mode is also malleable during decryption:

Flipping a bit in the ciphertext creates a random plaintext in the same block, but, as the ciphertext is XORed with the next plaintext block, bits at the same position in the next plaintext block are flipped accordingly.
Figure~\ref{figure:cbc_malleability} illustrates this malleability, using AES as the underlying cipher, where $k$ is the key of the cipher, $C_i$ is a ciphertext block (128 bits), $P_i$ the plaintext block (128 bits), and $IV$ the \acl{IV} of 128 bits.
Hence, XORing a $\Delta$ to the ciphertext $C_1$ leads to a random $P_1'$ instead of the correct plaintext $P_1$.
The plaintext in the next block $P_{2}$ is XORed with the $\Delta$ as well: $P_{2} \oplus \Delta$.
So by changing $C_i$, the attacker can flip arbitrary bits in $P_{i+1}$.

\subsection{Attack 1: Breaking Confidentiality}
\label{sec:attack_confidentiality}

The attack is essentially mounted in five steps:
\begin{compactenum}
\item Create a malicious bitstream and a readout bitstream
\item Configure the FPGA with the malicious bitstream
\item Reset the FPGA (automatically)
\item Read out the WBSTAR register using the readout bitstream
\item Reset the FPGA (manually)
\end{compactenum}
Using these five steps, the attacker can decrypt one word (32-bit) of the encrypted bitstream.
They can repeat these five steps for every word of the encrypted bitstream in order to recover it entirely. 


\newcommand\bfa{7}
\begin{figure*}
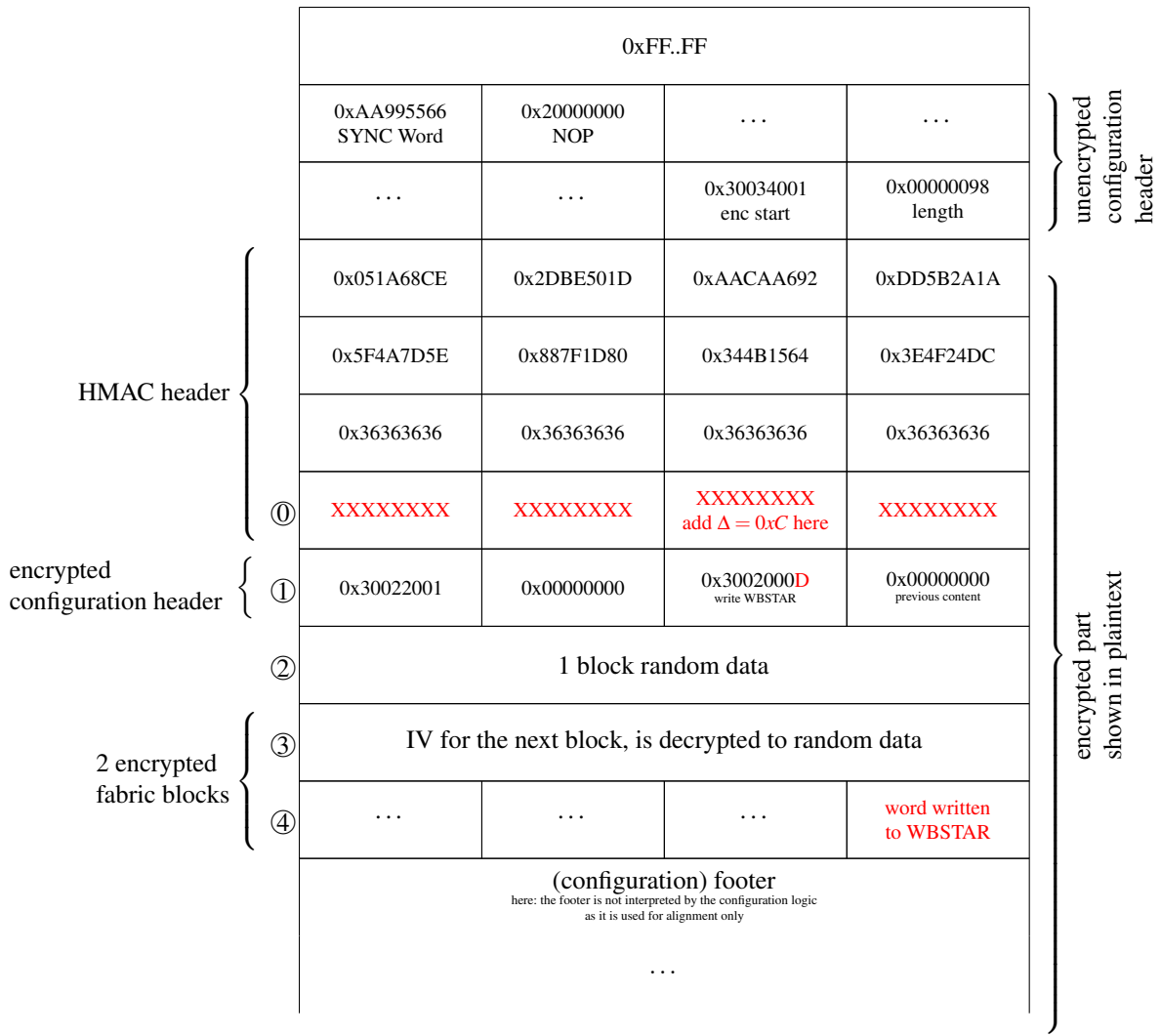

\centering
\begin{bytefield}[bitwidth=1.01em,bitheight=3em]{(4*\bfa+1)}
	\small
	\bitbox[]{1}{}&\bitbox{4*\bfa}{0xFF..FF} \\
	\begin{rightwordgroup}{\rotatebox{90}{\parbox{1.8cm}{unencrypted\\configuration\\header}}}
		\bitbox[]{1}{}&\bitbox{1*\bfa}{\footnotesize 0xAA995566\\{SYNC Word}} & \bitbox{1*\bfa}{\footnotesize 0x20000000\\NOP} & \bitbox{1*\bfa}{\dots} & \bitbox{1*\bfa}{\dots} \\
	
		\bitbox[]{1}{}&\bitbox{1*\bfa}{\dots} & \bitbox{1*\bfa}{\dots} & \bitbox{1*\bfa}{\footnotesize 0x30034001\\enc start} &\bitbox{1*\bfa}{\footnotesize 0x00000098\\length}
	\end{rightwordgroup} \\
	\begin{rightwordgroup}{\rotatebox{90}{\parbox{3cm}{encrypted part\\shown in plaintext}}}
		\begin{leftwordgroup}{HMAC header}
			\bitbox[]{1}{}&\bitbox{1*\bfa}{\footnotesize 0x051A68CE} & \bitbox{1*\bfa}{\footnotesize 0x2DBE501D} & \bitbox{1*\bfa}{\footnotesize 0xAACAA692} & \bitbox{1*\bfa}{\footnotesize 0xDD5B2A1A} \\
			\bitbox[]{1}{}&\bitbox{1*\bfa}{\footnotesize 0x5F4A7D5E} & \bitbox{1*\bfa}{\footnotesize 0x887F1D80} & \bitbox{1*\bfa}{\footnotesize 0x344B1564} & \bitbox{1*\bfa}{\footnotesize 0x3E4F24DC} \\
			\bitbox[]{1}{}&\bitbox{1*\bfa}{\footnotesize 0x36363636} & \bitbox{1*\bfa}{\footnotesize 0x36363636} & \bitbox{1*\bfa}{\footnotesize 0x36363636} & \bitbox{1*\bfa}{\footnotesize 0x36363636} \\
			\bitbox[]{1}{\circlearound{0}}&\bitbox{1*\bfa}{\footnotesize \color{red} XXXXXXXX} & \bitbox{1*\bfa}{\footnotesize \color{red} XXXXXXXX} & \bitbox{1*\bfa}{\footnotesize \color{red}  XXXXXXXX \\add $\Delta=0xC$ here} & \bitbox{1*\bfa}{\footnotesize \color{red} XXXXXXXX} 
		\end{leftwordgroup}\\
		
		\begin{leftwordgroup}{\parbox{3cm}{encrypted\\configuration header}}
			\bitbox[]{1}{\circlearound{1}}&\bitbox{1*\bfa}{\footnotesize 0x30022001} & \bitbox{1*\bfa}{\footnotesize 0x00000000} & 
			\bitbox{1*\bfa}{\footnotesize  0x3002000\textcolor{red}{D}\\\tiny write WBSTAR} & \bitbox{1*\bfa}{\footnotesize 0x00000000\\\tiny previous content} 
		\end{leftwordgroup}\\
		
		\bitbox[]{1}{\circlearound{2}} & \bitbox{4*\bfa}{1 block random data}\\

		\begin{leftwordgroup}{2 encrypted\\fabric blocks}
			\bitbox[]{1}{\circlearound{3}}&\bitbox{4*\bfa}{IV for the next block, is decrypted to random data}\\
			\bitbox[]{1}{\circlearound{4}}&\bitbox{1*\bfa}{\dots} & \bitbox{1*\bfa}{\dots} & \bitbox{1*\bfa}{\dots} & \bitbox{1*\bfa}{\footnotesize \color{red} word written\\to WBSTAR} 
		\end{leftwordgroup}\\

		\bitbox[]{1}{}&\bitbox[lrt]{4*\bfa}{(configuration) footer \\\tiny here: the footer is not interpreted by the configuration logic\\as it is used for alignment only}\\
		\bitbox[]{1}{}&\bitbox[lr]{4*\bfa}{$\dots$ } \\

		
	\end{rightwordgroup}
	
\end{bytefield}

	\caption{Attack bitstream  }
	\label{figure:bitstream_attack}
\end{figure*}

The malicious bitstream created in \textbf{Step 1} is shown in Figure~\ref{figure:bitstream_attack}, while a circled number corresponds to the block next to it; \circlearound{0} for instance corresponds to the last HMAC header block.
After the HMAC header, the previously initiated configuration header is completed as well, so boot configurations are written to the FPGA. 
In the first block~\circlearound{1} (the last two words) already, the command (0x30020001) writes one word to the WBSTAR register.
In a default configuration, the content written to this register is zero, but it is irrelevant for the attack. 
In more detail, the first bits of the command 0x3 issues a type 1 package with a write operation, while the 0x020 points to the WBSTAR register. 
The 0x01 at the end states the written length in words, here a single word. 
Thus, the following word, 0x00000000, is written to the WBSTAR register. 
The bitstream generation in Vivado is deterministic, meaning that the commands in the encrypted header are the same among different bitstreams and change only in their configuration content, i.e., the words after the commands. 
Thus, the adversary can assume the plaintext in the encrypted header, so they can, for instance, know the position of the write WBSTAR command. 
We here chose the WBSTAR command, because it is the last command in the first block after the HMAC header, but any other command in the first block could be chosen.

During a reset, the whole fabric and the configuration registers are set to their default values (mostly zero), but crucially, the WBSTAR (warm boot start address) and BOOTSTS (boot status) registers are \emph{not} reset.
These registers are used for the MultiBoot and fallback feature, which enables the FPGA to boot from a different SPI/BPI memory address to safely update the FPGA or boot designs with different functionality.
This MultiBoot feature uses the content of the WBSTAR register as the boot address for the attached non-volatile memory.
Hence, the WBSTAR register cannot be cleared during a reset, as its address might be needed for the boot process.

The attack is also based on the fact that when writing more than one word to a single word register, only the last written word is stored in that register. 
For example, if the bitstream in Figure~\ref{figure:bitstream_structure} is changed such that the length of the WBSTAR write operation is 4 (not 1), 
then the four words after the command would be written to the WBSTAR register, but only the last word is finally stored there, i.e., the NOP command 0x20000000. 
The number of written words can even extend further, so that configuration data from the fabric, which comes later in the bitstream, is written to the WBSTAR register. 

Now the attacker can change this length field in the encrypted bitstream by adding a $\Delta$ to the corresponding word of the write WBSTAR instruction of the former block \circlearound{0}, thereby exploiting the CBC malleability.
Consequently, the manipulated block (the HMAC header) becomes random data (marked with X in Figure~\ref{figure:bitstream_attack}).
However, this change is irrelevant for the bitstream, as the other changes will result in a faulty HMAC validation anyway and the bitstream is still valid for the configuration logic, since only the HMAC ipad is changed.

The length of the encrypted data, without the HMAC header and footer, must be a multiple of 512 bits as the SHA-2 operates on 512-bit chunks~\cite{moradi2011vulnerability}. 
Hence, the encrypted data must be at least four AES blocks long, so we set the length of the write WBSTAR operation to $0xD = 13$ ($\Delta = 0xC$), i.e., all data until the end of the, first and only, 512-bit chunk of encrypted data are written to the WBSTAR register. 
At this position, i.e., at the end of the 512-bit chunk, we place the ``to be decrypted'' block \circlearound{4}. 
This block can be any AES block from the encrypted bitstream, which includes the encrypted fabric data. 
Since the bitstream encryption uses the CBC mode, the former block before \circlearound{3} must be the block from the CBC chain, i.e., it is the IV for the decrypted block. 
An additional block \circlearound{2} is needed to fill the 512-bit chunk (which can be random). 
It is placed between the block of the WBSTAR write operation and the two decryption blocks.
Note that this random block \circlearound{2} and the IV block \circlearound{3} are decrypted to random data. 
Since the WBSTAR write operation writes 13 words, all random data is interpreted as data which are stored in the WBSTAR register,
but only the desired decrypted word is stored in the WBSTAR register.

To readout the other words in the last blocks \circlearound{4}, the $\Delta$ should be changed accordingly, i.e., the write length is set to $10$, $11$, or $12$.
However, the configuration logic will interpret the last (next to last, ...) words as a standard package, which might be data from the fabric. 
Thus, they are not correct instructions, and they might cause unwanted random commands. 
Hence, a second $\Delta$ is added to the IV block to change the 13th (and 12th, 11th) word to a NOP command, which is possible as the attacker first decrypts the last block and uses the CBC malleability again.
This prevents the configuration logic from falsely interpreting the last words.

Next, in \textbf{Step 2}, the FPGA is configured with the malicious bitstream. 
Due to the changes made to the bitstream before, the HMAC is invalid.
The configuration logic correctly detects this and resets the FPGA in \textbf{Step 3} automatically. 
Nevertheless, the HMAC is only checked at the end.
Thus, the WBSTAR register has already been written before the check failed.

In \textbf{Step 4}, a (not encrypted) readout bitstream is sent to the FPGA to obtain the content of the WBSTAR register.
This bitstream is not encrypted as no interaction with the fabric is made.
It reveals the one word written to the register.
The full readout bitstream can be obtained from Appendix~\ref{sec:readoutBS}.

Lastly, in \textbf{Step 5}, the FPGA is reset manually to repeat the steps.
Otherwise, multiple readouts would fail.
On the JTAG interface, the JPROGRAM command is sent, and on the SelectMAP interface, the PROGRAM\_B pin is pulsed low to issue the reset. 
This clears the fabric memory and is sufficient to reset the configuration logic as well. 


\subsection{Attack 2: Breaking Authenticity}
With the first attack, the FPGA can be used to decrypt arbitrary blocks.
Hence, it can also be seen as a decryption oracle. 
Thus, we can also use this oracle to encrypt a bitstream, as shown by Rizzo and Duong in~\cite{DBLP:conf/woot/RizzoD10}, and generate a valid HMAC tag. 
Let $dec_{KAES}(\cdot)$ be the decryption function of the target FPGA configured with the AES key $KAES$, $C_i$ a ciphertext block, and $P_i$ the corresponding plaintext block following the underlying CBC mode.
Therefore, it holds (CBC function),
\begin{align}
P_i = dec_{KAES}(C_i) \oplus C_{i-1}.
\end{align}
Suppose, $C_i$ and $C_{i-1}$ are arbitrarily selected. 
We can use the FPGA as the decryption oracle and find out $P_i$, with using the introduced attack in Section~\ref{sec:attack_confidentiality},

The goal is to find $C_i'$, which generates the desired $P_i'$ inside the FPGA.
To this end, we just need to set (CBC malleability)
\begin{align}
C_{i-1}' = P_i \oplus C_{i-1} \oplus P_i'.
\end{align}
For the previous block $P_{i-1}'$ we can find (for an arbitrary selected $C_{i-2}$),
\begin{align}
 P_{i-1} = dec_{KAES}(C_{i-1}') \oplus C_{i-2}
\end{align}
while using the FPGA as the decryption oracle again.
Similarly, we can set 
\begin{align}
 C_{i-2}' = P_{i-1} \oplus C'_{i-2} \oplus P_{i-1}'
\end{align} which leads to generate the desired plaintext block $P_{i-1}'$. 
This process is repeated toward the first block $P_1'$, and the IV is set to $C_0'$ in the unencrypted header.

Therefore the attacker can encrypt an arbitrary bitstream by means of the FPGA as a decryption oracle. 
The valid HMAC tag can also be created by the attacker, as the HMAC key is part of the encrypted bitstream.
Hence, the attacker can set his own HMAC key inside the encrypted bitstream and calculate the corresponding valid tag. 
Thus, the attacker is capable of creating a valid encrypted bitstream, meaning the authenticity of the bitstream is broken as well.

\subsection{Wrap-Up: What Went Wrong?}

These two attacks show again that nowadays, cryptographic primitives hold their security assumptions, but their embedding in a real-world protocol is often a pitfall.
Two issues lead to the success of our attacks: First, the decrypted data are interpreted by the configuration logic before the HMAC validates them. 
Generally, a malicious bitstream crafted by the attacker is checked at the end of the bitstream, which would prevent an altered bitstream content from running on the fabric. 
Nevertheless, the attack runs only inside the configuration logic, where the command execution is not secured by the HMAC.

Second, the HMAC key $K_{HMAC}$ is stored inside the encrypted bitstream. 
Hence, an attacker who can circumvent the encryption mechanism can read $K_{HMAC}$ and thus calculate the HMAC tag for a modified bitstream. 
Further, they can change $K_{HMAC}$, as the security of the key depends solely on the confidentiality of the bitstream.
The HMAC key is not secured by other means.
Therefore, an attacker who can circumvent the encryption mechanism can also bypass the HMAC validation.

\section{Case Studies}
\label{sec:caseStudys}
We conducted several experiments to validated the attacks. 
We tested the attacks on the Xilinx Kintex-7 (XC7K160T), mounted on a SAKURA-X Board~\cite{sakurax},  on a Xilinx Artix-7 (XC7A35T), mounted on a Basys3 board~\cite{Basys3}, and on a Xilinx Virtex-6 (XC6VLX240T), mounted on the ML605 evaluation kit. 
Since the Xilinx user guid~\cite{ug7Config} states no difference between the 7-Series configurations engines, we conclude that our attack is applicable to all 7-Series devices.
\todo[inline]{ask xilinx and claim more}
We first attacked the SelectMAP interface on the Kintex-7.
For this, we implemented a controller on the Spartan-6 FPGA, which is mounted aside the Kintex-7 on the SAKURA-X board.
The Spartan-6 can configure the Kintex-7 via the SelectMAP interface.
The controller on the Spartan-6 and a controlling computer are connected via UART.
The computer sends the bitstream to the controller, where it is saved in a BRAM and is transmitted to the  Kintex-7 under attack. 

After the first successful attack, we also implemented the attack on the Basys3 board.
Here, we used the open-source xc3sprog~\cite{xc3sprog} to configure the Artix-7 via the onboard USB programmer.
In order to validate that an adversary can use the JTAG interface, we implemented the attack for the JTAG interface with a SEGGER J-Link EDU~\cite{JLink}.
We used OpenOCD~\cite{OpenOCD} to utilize the J-Link and used the scripting engine of OpenOCD to pass the individual bitstream's bytes to the JTAG interface. 
Since there is a lot of static data, i.e., only two fabric blocks change per 32-bit readout, and the USB interface to the J-Link is slow, we implemented the attack on a microcontroller. 

We used the STM32F407G-DISC1 discovery kit~\cite{STM32Disc} equipped with an STM32 microcontroller.
It emulates a JTAG controller and is connected via a UART to a controlling PC.
The microcontroller retrieves the encrypted header only once, while it gets large chunks of the bitstream and sends it to the FPGA.
It individually adds $\Delta$ to the encrypted header to increase the performance as no roundtrip from and to the PC is needed. 
The microcontroller itself generates the JTAG clock.
Inbetween the JTAG clock tick, a single bit is put on an I/O pin.
Thus, there are multiple instructions on the microcontroller to transmit a single bit, i.e., at least set a data bit, reset the clock pin, set the clock pin.
Hence, there might be performance improvement possible.
Note that every readout needs two small bitstreams to be loaded.
First, the malicious bitstream to write the WBSTAR register is transmitted.
This bitstream is 211~words long.
Second, the readout bitstream is sent to the FPGA, which is 22~words long.
Additionally, the FPGA resets two times.
Using the implementation of the microcontroller over the JTAG interface, a readout of a single 32-bit word is done in 7.9 milliseconds.
Since the XC7K160T's bitstream has a size of 53,540,576~bits, the readout of the bitstream completes in 3 hours and 42 minutes. 
Even with the largest 7K480T with a bitstream size of 149,880,032~bits, the attack can run in approximately 10~hours. 
In Table~\ref{tab:runtime}, we  selected some FPGAs as examples and provided estimated runtimes of the attack. 
Note that the attack can also be parallelized if two FPGAs with the same bitstream are available, which can be the case, e.g.,  if one global key is used for bitstream encryption within a given product. 
Even though less desirable from a security point of view, using a global key, is without doubt, a tempting option in many real-world situations as it dramatically simplifies key management. 

\begin{table}
\begin{center}
\begin{tabular}{lrr}
\textbf{FPGA} & \textbf{Bitstream Size (Bits)} & \textbf{Time (HH:MM)}\\\hline
\hline
7S6 & 4310752 & 00:18\\\hline
7S50 & 17536096 & 01:12\\\hline
7S100 & 29494496 & 02:01\\\hline
\hline
7A12T & 9934432 & 00:41\\\hline
7A35T & 17536096 & 01:12\\\hline
7A200T & 77845216 & 05:20\\\hline
\hline
7K70T & 24090592 & 01:39\\\hline
7K160T* & 53540576 & 03:42\\\hline
7K480T & 149880032 & 10:17\\\hline
\hline
7VX1140T & 385127680 & 26:25\\
\end{tabular}
\caption{Expected runtime of the attack on various Xilinx FPGAs, (*) extrapolated from the XC7K160T}
\label{tab:runtime}
\end{center}
\end{table}

Running the second attack on the authentication of the bitstream requires the same amount of time as the first attack.
Because within the second attack the whole bitstream is encrypted.
We can even speed up the second attack when only parts of the bitstream need to be altered, e.g., when the attacker wants to introduce a hardware Trojan and only changes to a small fraction of the design are required.
We also note that no design can fully utilize the entire FPGA.
Consequently, there are blocks of the bitstream that are unused by the design. 
If now an attacker re-encrypts the changed blocks of the design, until they reach an unused block, they can stop the re-encryption and utilize this unused block as the IV for the next regular block (CBC malleability).
Consequently, the unused block will be decrypted to random data.
However, since its content is not necessary for the design, it can be ignored. 
Thus the attacker only needs to re-encrypt a part of the bitstream rather than the whole bitstream, which speeds up the encryption process.

Furthermore, we evaluated old FPGA series. 
The configuration logic on Virtex-6 devices is mostly identical with the 7-Series' configuration logic. 
Hence, we also mount our attack on the XC6VLX240T, using xc3sprog with the USB JTAG port present on the ML605 development board. 
The single shortcoming of the attack is the limitation of the WBSTAR register. 
The start address, present in the WBSTAR, is shortened by 3-bit compared to the 7-Series. 
Hence, the upper 3 bits are marked as reserved.
But only the two leftmost bits are not implemented, i.e., writing any arbitrary value to those 2 bits will always return zero. 

Therefore, every upper 2 bits of all words in the bitstream cannot be read out, which leads to an imperfect recovered netlist. 
Imperfect netlists are an already known obstacle in the reverse-engineering community and can be tackled to a certain degree~\cite{DBLP:journals/tifs/ErozanHBAT20}. 
Moreover, the encoding of the PIPs in the bitstream and a meaningful routing of nets can help to repair the recovered netlist \cite{ender2019insights}. 
However, the reversed LUTs are not unambiguously recoverable. 

\section{Countermeasures \& Defense Techniques} 
\label{sec:countermeasures}
In this section, we discuss two possible countermeasures and four defense techniques. 
We define countermeasures as techniques defending current 7-Series devices, e.g., which hardware developers can use, and defense techniques as measures, which require to update the silicon, e.g., which platform companies like Xilinx can offer.
Note that our attacks are based on protocol flaws that are hard-coded in the FPGA silicon.
Thus, any kind of non-trivial change to the security protocol is not possible without a re-design of the FPGA hardware and is currently not available for 7-Series and Virtex-6 devices.
Table~\ref{tab:countermeasure} gives an overview of our proposed defence techniques and countermeasures, which are discussed in this section.
We divided the section into two parts.
In the first part, we discuss four defense techniques for new developments, while the first two are (seemingly) implemented in the new Xilinx series. 
In the second part, we discuss about design obfuscation and a patch to the PCB as raise-the-bar countermeasures for current 7-Series devices.


\begin{table}
 \begin{center}
\begin{tabularx}{0.49\textwidth}{X|>{\centering\arraybackslash}p{1.1cm}|>{\centering\arraybackslash}p{1.1cm}|>{\centering\arraybackslash}p{1.15cm}}
    Section & new dev & new series & current 7-Series\\\hline
	\ref{sec:validateBeforeUse} Validate before use & $\bullet$ & $\bullet$ &\\\hline
	\ref{sec:patchableEnc} Patchable Enc & $\bullet$  & $\bullet$ &\\\hline
	\ref{sec:IFA} IFA, Model Checker & $\bullet$  & &\\\hline
	\ref{sec:opensource} OpenSource HW & $\bullet$  & &\\\hline
	\ref{sec:obfuscation} Obfuscation & & & $\bullet$ \\\hline
	\ref{sec:RSPin} RS pin reset & & & $\bullet$
 \end{tabularx}
 \end{center}
\caption{Proposed countermeasures and defense techniques and their adaptability on new developments, new series from Xilinx, and the current 7-Series discussed in Section~\ref{sec:countermeasures}}
\label{tab:countermeasure}
\end{table}

\subsection{General Defense Techniques}
Here we discuss on general defense techniques, which can be offered by Xilinx and are already partially used in the new UltraScale(+) and Zynq series. 

\subsubsection{Validate Before Use}
\label{sec:validateBeforeUse}

In a sound security design, no data is interpreted before its cryptographic validation. However, one of the root causes of the decryption attack is that this principle is violated in the FPGA's encryption engine, i.e., data of the encrypted bitstream header is interpreted before it has been verified. 
Hence, the apparent countermeasure is to validate the configuration header before any action. If that could be implemented, the attack would be detected as it manipulates the header. 
Nevertheless, to our knowledge, updating the bitstream encryption engine on current devices is not possible, as it is implemented in the silicon and would require a redesign. 

It is instructive to look at the newer FPGA families by Xilinx. It seems that Xilinx introduced a continuous checksum in the UltraScale and UltraScale+ series, as we could not mount the attack on such devices. 
Xilinx used an AES-GCM scheme for the new series, where the first 32 bits of every 256-bit encrypted data block are unknown (seems random), which are also not addressed by the configuration logic.
We speculate that these 32 bits are  a kind of checksum used for verification/integrity.
However, to the best of our knowledge, there is no official statement from Xilinx about these 32 bits.

\subsubsection{Patchable Bitstream Encryption} \label{sec:patchableEnc}
It might be a bit ironic that the security measures of a reconfigurable device are not reconfigurable.
Unterstein et al. showed an implementation of a patchable bitstream encryption scheme on the Zynq-7000 platform~\cite{Unterstein:2019:SSU:3338508.3359573}, which is a realization of~\cite{Zynq7000Authentication}.
There are several variations of the same idea reported in references~\cite{8226019, owen2018autonomous, Kashyap:2016:COS:2854101.2816822}.
Note that the Zynq-7000, UltraScale and UltraScale+ devices have the needed public key scheme, while the 7-Series and older devices have not. 
Consequently, this countermeasure does not apply to the 7-Series.
In a nutshell, the FPGA loads an initial bitstream and only verifies it.
This initial bitstream contains a hardened bitstream encryption engine in terms of side-channel resistance, and a \ac{PUF} which generates the encryption key.
This engine decrypts the original bitstream, as well as loads it via partial-reconfiguration to the fabric.

The engine is patchable as it residents in the fabric and is not hard-coded into the FPGA.
Hence, it is possible to improve the engine if new attacks arise, e.g., enhanced side-channel attacks, or if bugs are found in the system (like our attacks in this work).
Additionally, no key storage is needed as a \ac{PUF} is used, which reduces the risk of attacks and disadvantages of the BBRAM and eFUSES key storage implemented by Xilinx.
For example, it is shown in~\cite{lohrke2018key,trimberger2014fpga} how to read out keys stored in register cells of various FPGAs.

The only requirement is to verify the initial bitstream and avoid running invalid bitstreams.
Accordingly, the authors of~\cite{Unterstein:2019:SSU:3338508.3359573} used the Xilinx Zynq-7000 series, where a public-key signature scheme is integrated. 
Besides the Zynq-7000 series, the UltraScale and UltraScale+ series also include such embedded public-key signatures. 
In the 7-Series devices, no such signature scheme exists.
The validating prevents any modification, e.g., the insertion of hardware Trojans like modifications to the encryption engine to leak the keys or the decrypted bitstream.
Since the initial bitstream only needs to be verified, it is not encrypted.
Hence, the attacker can see how it is realized and implemented.
Thus, the implementation and exact location of the \ac{PUF} is known to the adversary.
Therefore, the FPGA must suppress any non-verified bitstreams, as otherwise, the attacker can modify the bitstream to read out the PUF response, i.e., the secrets.

Admittedly, this scheme is based on trust in the public key signature scheme and its implementation. 
Although it lowers the unpatchable attack surface to the signature scheme only, as if a successful attack targets the encryption scheme, it is still patchable.
However, an unpatchable attack surface exists. 
Thus, we discuss model-checking and \ac{IFA} as another countermeasure in the following.

\subsubsection{\acl{IFA} and Model Checking} \label{sec:IFA}
A detailed study of the Xilinx official documents~\cite{ug7Config}, together with experiments, led us to our attack. However, since the bitstream encryption and the behavior of the WBSTAR register are documented, it is perceivable that one could have developed a formal model to find the bug.  Within the last years, there has been an increasing trend towards formal verification and model checking in the scientific community.
The recent publication from Dessouky~\cite{dessouky2019hardfails} discusses various techniques to find hardware bugs.
Three of them can be applied to our findings: proof assistant and theorem-proving, model checking, and \ac{IFA}.
Note that the formal verification of the design against the specification is not sufficient, as the bug is already visible in the documentation, i.e., the specification of configuration. 
Additionally, after the specification is changed, the current devices should be reproduced to apply this countermeasure.

Within proof assistant and theorem-proving, the security properties are mathematically modeled and verified with the proofs.
For example, VeriCoq~\cite{bidmeshki2015vericoq} transfers a Verilog code into the Coq language and proof system.
With additional labeling the signals, the flow of information is tracked, i.e., the signals are classified if they transmit secret information or not.
Mathematical proofs ensure that no secret information is leaked. 
However, the accurate labeling of each signal is error-prone and laborious, and the proving might be infeasible for large designs.

The more general model checking is mostly built on Boolean satisfiability problems.
The engineer formulates an abstract model of the specification and tests predefined assumptions of the model to be correct, e.g., a decrypted bitstream cannot flow to a configuration register.

Since model checkers are a general approach and require to write an additional model besides the specification and HDL code, \ac{IFA} checks the design directly.
In general, the input data are labeled, mostly in high and low, e.g., private and public information.
Then, these labels are tracked while flowing through the design.
If any private-labeled data influence public data, a vulnerability might be detected. 
For performing \ac{IFA}, a variety of tools exists which operate at different layers of abstraction.
\ac{GLIFT} works on the gate-level~\cite{ardeshiricham2017register,oberg2010theoretical,tiwari2009complete}, where the analysis is performed on the synthesized design.
Hence, it is mostly done automated and works on existent designs but does not scale well.
Caisson~\cite{li2011caisson}, Sapper~\cite{li2014sapper}, and SecVerilog~\cite{zhang2015hardware} works on the language level, while Caisson and Sapper are new HDLs, and SecVerilog extends Verilog with annotations.
Therefore, it is applicable to already existing projects.

\subsubsection{Open-Source Hardware} \label{sec:opensource}
When considering a redesign, one can take open-source hardware into account.
Open-source hardware has, at the least in theory, the advantage of being verifiable from a large community, similar to what is already done in software projects, e.g., OpenSSL~\cite{beringer2015verified}.
Hence, it gains its trust by transparency rather than obscurity and follows the approach of Kerckhoffs Principle~\cite{hoepman2008increased}.
The recently released OpenTitan~\cite{openTitan, openTitanGithub} silicon root of trust moves into that direction. 
It provides a trust anchor for system security and is applicable as an IP core for custom made devices.

\subsection{Countermeasures for Current Devices}
With our attack, a product using a 7-Series (or Virtex-6) device needs to be upgraded to one with a sound bitstream encryption engine, as our findings imply a complete loss of authenticity and confidentiality and no patch is available. \todo{ask xilinx: available vs possible}
However, it is neither possible nor feasible to update the FPGAs used in all products.
The old Virtex-6 and current 7-Series are commonly used in low-budget devices.
Thus, a countermeasure that raises the bar for the attacker can be sufficient in many applications.
In this section, we first introduce obfuscation as a countermeasure and then a patch to the PCB to reset the FPGA if the attack is detected. 

\subsubsection{Obfuscation} \label{sec:obfuscation}
One of the raise-the-bar countermeasures is obfuscation.
It changes the design without changing its functionality, while the design is concealed and becomes significantly more complex to be reverse-engineered for humans and machines.
Several works~\cite{goren2013partial, fyrbiak2018difficulty, porter2009dynamic, goren2011fpga} already exist, especially for low-budget FPGAs, which do not offer any bitstream encryption.
Here, two main goals exist: securing against overproduction/cloning and reverse engineering.
A mechanism to secure against overproduction is always bound to an FPGA, hence often \acp{PUF} are used.
Based on physical variations of each device, a device-specific key is generated by the PUF to unlock the design. 
Consequently, the design would not unlock on a different FPGA, since its physical characteristics are different, and the PUF generates a different unlock key. 
More general obfuscation schemes defend against reverse-engineering.
Often, the \acp{FSM} of the designs are targeted; hence, dummy states are added to the design to increase the complexity of the state-transition graph. By applying a transition sequence to unlock the design, the \ac{FSM} can still be used as initially intended.
Subramanyan et al. benchmarked different obfuscation techniques in~\cite{DBLP:conf/host/SubramanyanRM15}.
They consider an area overhead of 5\% as a realistic budget and a 10\% overhead for sensitive designs acceptable. 
Nevertheless, the current obfuscation methods are not ideal, as shown in~\cite{fyrbiak2018difficulty}.


\subsubsection{Revision Select PIN}\label{sec:RSPin}
In the second raise-the-bar countermeasure, the \ac{RS} pins are used to reset the FPGA and clear the BBRAM key storage, which extends the root of trust from the FPGA's silicon to the PCB. 

\todo[inline]{root-of-trust zu stark? maybe trusted computing base }

Besides the warm boot address (bits 28-0), the WBSTAR register drives two \ac{RS} pins (bit 31 and 30) during the configuration phase.
The two RS pins are enabled with the \emph{RS\_TS\_B} bit in the WBSTAR register (bit 29), which controls a tri-state, driving the RS pins.
If the RS\_TS\_B bit is high, these two RS bits in the WBSTAR register directly drive the two RS pins.
Otherwise, the RS pins are in high-Z.
During regular operation, e.g., after the configuration phase, the RS pins can be used as regular I/O pins~\cite{ug7Config}.
We have observed this on the SAKURA-X board, where one RS pin drives one of the user's LEDs.

During the attack, the bitstream content is written into the WBSTAR register.
Thus, the RS pins are driven with the bitstream contents.
Exemplarily, if the RS\_TS\_B bit is set in any word, the two RS bits in that word drive the RS pins. 
In a raise-the-bar countermeasure, one could wire the RS pins to a reset logic on the PCB, to power-off the FPGA, which hinders the readout of the current word as the power-down wipes all registers, including the WBSTAR register.
Accordingly, if the upper three bits of a random word in the bitstream are set, the FPGA would be reset, which hinders the readout of the WBSTAR content.
Nonetheless, a defender needs to impede the readout of further bitstream words, since words where no RS and the RS\_TS\_B bit is set, are still possible to be read out as the FPGA is not reset during the attack.
Therefore, one can also cut the battery power to the BBRAM to clear the stored key.
Thus, no further encryption operations are possible since the keys are discarded.
To provoke the reset via an RS pin, the defender can change the bitstream content in unused regions to drive an RS pin high, i.e., set the upper 3 bits in multiple words of the bitstream that are unused.

On the PCB, the RS pins are wired to a reset circuit to power-cycle the FPGA completely, i.e., the FPGA is reset, and the battery power to the  BBRAM is cut.
Whereas, the bitstream encryption's goal is to run all security-relevant measures inside the FPGA and not rely on other components.
This means that the FPGA's designer needs to solely trust the FPGA's silicon and not other components, which minimizes the attack surface notably.
Thus, with this countermeasure, the bitstream encryption's goal of not relying on the PCB is not fulfilled. Hence, this method is a raise-the-bar countermeasure.

\section{Conclusion}
\label{sec:conclusion}
In this paper, we demonstrated two attacks on the Xilinx 7-Series and Virtex-6 bitstream encryption. 
The first attack breaks the confidentiality of any encrypted design using the FPGA as a decryption oracle. 
The second attack breaks the authenticity by using the same oracle to encrypt arbitrary bitstreams and generating a valid authentication tag. 
In our implementation, any communication with the oracle needs 7.9\,ms to reveal 32 bits of an encrypted block.
Thus, it takes for example 3:42 hours to recover a Kintex-7 XC7K160T bitstream (see Table~\ref{tab:runtime}).

For our  attacks, it is sufficient to have access to the encrypted bitstream and either the JTAG or the SelectMap configuration interface.
Hence, the attack can be potentially conducted remotely and does not require any sophisticated tools.
We identified two roots leading to the attacks. First, the decrypted bitstream data are interpreted by the configuration logic before the HMAC validates them.
Second, the HMAC key is stored inside the encrypted bitstream.
Consequently, if the confidentiality is broken, the authenticity is lost as well.

We consider this as a severe attack, since (ironically) there is no opportunity to patch the underlying silicon of the cryptographic protocol.
We note that the 7-Series have a substantial share of the FPGA market, which makes it even more difficult or impossible to replace these devices.
As a countermeasure, we propose (for future-series devices) to verify all input data before use, apply model checkers and \ac{IFA}, use when possible open-source hardware, and make use of a patchable bitstream encryption engine, like the one implemented on the Zynq-7000.
For the current series, we propose to use obfuscation schemes or patching the PCB to use the FPGA's RS pins for clearing the BBRAM key storage in case of an attack.
Although these countermeasures are not a substitute for a sound bitstream encryption, they still raise the bar for legacy systems until more secure devices can be provided.

The bitstream encryption for newer generations, e.g., UltaScale, appears to be an entirely new development.
Thus, it is still impossible to mount the same attacks on new-generation devices, as detailed information about the bitstream packets is not yet publicly available.

\todo[inline]{Bildunterschriften}

\section*{Acknowledgments}
We communicate these findings with Xilinx in a responsible disclosure on 24 September 2019. We would like to thank Xilinx for their support and kind communication during this process, as well as the anonymous reviewers and our shepherd Stefan Mangard for all their helpful comments.
Part of this work was supported by the European Research Council (ERC) under the European Union’s Horizon 2020 Research and Innovation programme (ERC Advanced Grant No. 695022 (EPoCH)), as well as, by the German Research Foundation (DFG) within
 the framework of the Excellence Strategy of the Federal Government and
 the States - EXC 2092 CASA - 390781972.

\todo[inline]{SPI leak WBSTAR via address bits?}

\todo[inline,color=green]{Either provide experimental evidence, or significantly scale back, the claims about the ability to run this attack remotely, including from the abstract
}

\todo[inline,color=green]{older FPGAs, outside the Kintex 7 and Artix 7 series.
}

\todo[inline,color=green]{Discuss in more detail the requirements of the attack, both physical (connections, interfaces, etc) and logical (knowledge of the plaintext, how is Delta calculated, etc)
}

\todo[inline]{Clarify the "unpatchable" claims, incorporating feedback from Xilinx
}

\todo[inline]{Improve the presentability of the paper by correcting typos and other minor items, as highlighted by the reviewers
}

\todo[inline,color=green]{The description of the data format and encryption/integrity mechanism that is analyzed needs to be significantly improved. See comments from reviewer}

\bibliographystyle{plain}
\bibliography{refs}

\appendix
\bigskip
\section{Readout Bitstream}
\label{sec:readoutBS}
\begin{lstlisting}[caption=Readout Bitstream]
0xFF, 0xFF, 0xFF, 0xFF, 
0xFF, 0xFF, 0xFF, 0xFF, 
0xFF, 0xFF, 0xFF, 0xFF, 
0xFF, 0xFF, 0xFF, 0xFF, 
0xFF, 0xFF, 0xFF, 0xFF, 
0xFF, 0xFF, 0xFF, 0xFF, 
0x00, 0x00, 0x00, 0xBB, 
0x11, 0x22, 0x00, 0x44, #BUS Size Detect
0xFF, 0xFF, 0xFF, 0xFF, 
0xFF, 0xFF, 0xFF, 0xFF, 
0xAA, 0x99, 0x55, 0x66, #SYNC Word
0x20, 0x00, 0x00, 0x00, #NOP
0x30, 0x00, 0x80, 0x01, 
0x00, 0x00, 0x00, 0x04, 
0x20, 0x00, 0x00, 0x00, 
0x20, 0x00, 0x00, 0x00, 
0x20, 0x00, 0x00, 0x00, 
0x28, 0x02, 0x00, 0x01, #read reg WBSTA
0x20, 0x00, 0x00, 0x00, 
0x20, 0x00, 0x00, 0x00, 
0x20, 0x00, 0x00, 0x00,
0x20, 0x00, 0x00, 0x00
\end{lstlisting}

\listoftodos

\end{document}